\documentclass[twocolumn]{aastex63}
\usepackage{geometry}
\usepackage{amsmath}
\usepackage{graphicx}
\usepackage{color}

\usepackage{tikz}
\usetikzlibrary{matrix}
\usepackage{natbib}

\usepackage{lineno}

\graphicspath{{./}{Figures/}}

\begin{document}


\title{Are Delayed Radio Flares Common in Tidal Disruption Events?\\[1ex] The Case of the TDE iPTF\,16fnl}

\author[0000-0002-5936-1156]{Assaf Horesh}
\affiliation{Racah Institute of Physics, The Hebrew University of Jerusalem, Jerusalem 91904, Israel}
\author[0000-0003-0466-3779]{Itai Sfaradi}
\affiliation{Racah Institute of Physics, The Hebrew University of Jerusalem, Jerusalem 91904, Israel}
\author{Rob Fender}
\affiliation{Astrophysics, Department of Physics, University of Oxford, Keble Road, Oxford, OX1 3RH, UK}
\author{David A. Green}
\affiliation{Astrophysics Group, Cavendish Laboratory, 19 J. J. Thomson Ave., Cambridge CB3 0HE, UK}
\author{David R. A. Williams}
\affiliation{Jodrell Bank Centre for Astrophysics, School of Physics and Astronomy, The University of Manchester, Manchester, M13 9PL, UK}
\affiliation{Astrophysics, Department of Physics, University of Oxford, Keble Road, Oxford, OX1 3RH, UK}
\author{Joe Bright}
\affiliation{Astrophysics, Department of Physics, University of Oxford, Keble Road, Oxford, OX1 3RH, UK}

\begin{abstract}

Radio emission from tidal disruption events (TDEs) originates from an interaction of an outflow with the super-massive black hole (SMBH) circum nuclear material (CNM). In turn, this radio emission can be used to probe properties of both the outflow launched at the event and the CNM. Until recently, radio emission was detected only for a relatively small number of events. While the observed radio emission pointed to either relativistic or sub-relativistic outflows of different nature, it also indicated that the outflow has been launched shortly after the stellar disruption. Recently, however, delayed radio flares, several months and years after stellar disruption, were reported in the case of the TDE ASASSN-15oi. These delayed flares suggest a delay in the launching of outflows and thus may provide new insights into SMBH accretion physics. Here, we present a new radio dataset of another TDE, iPTF\,16fnl, and discuss the possibility that a delayed radio flare has been observed also in this case, $\sim 5$\,months after optical discovery, suggesting that this phenomenon may be common in TDEs. Unlike ASASSN-15oi, the data for iPTF\,16fnl is sparse and the delayed radio flare can be explained by several alternative models: among them are a complex varying CNM density structure and a delayed outflow ejection. 

\end{abstract}

\section{Introduction}
\label{sec:intro}

Tidal disruption events (TDEs) occur when a star is tidally disrupted by a super-massive black hole (SMBH) when passing close enough to the SMBH \citep{rees_1988}. Tens of electromagnetic transients (either at high energy or at optical/UV bands) observed in galaxy nuclei over the last several decades are interpreted as originating from TDEs. The physical process responsible for the transient emission is still highly debated. Whether it is a result of an accretion disk formed from the bound stellar debris (\citealt{rees_1988,Phinney_1989,evans_1989}), self-intersecting streams of stellar debris \citealt{shiokawa_2015,Bonnerot_2020, piran_2015}), or some other process remains to be seen. 

TDEs can also lead to the launching of high-velocity outflows. These outflows can be, for example, jets that are driven by accretion, accretion disk wind outflows, or even unbound stellar debris traveling away from the SMBH. The energetics and temporal behaviour of an outflow launched in a TDE may vary based on its origin. One of the main ways to detect and monitor a wide variety of outflows is via radio observations. The interaction of a high-velocity outflow with the SMBH circum nuclear material (CNM) will drive a shockwave into the CNM that in turn will lead to synchrotron emission. Radio observation of this synchrotron emission can be used to measure the shockwave (and hence the outflow) velocity, measure the magnetic field strength at the shock front, and to also measure the CNM density. So far, however, radio emission has been detected in a relatively small number of events. In the most prominent high-energy event to-date, {\it Swift}\,J1644+57 (\citealt{Bloom_2011,Burrows_2011}), bright radio emission revealed a relativistic jet (\citealt{Zauderer_2011,Berger_2012}) and provided calorimetric measurements for over a decade. However, in a large sample of optically-discovered TDEs, such radio emission has not been observed (see a review by \citealt{Alexander_2020}). Only in a handful of optical TDEs, radio emission has been detected but with a luminosity orders of magnitudes lower than in {\it Swift}\,J1644+57. In this handful of events, analysis of the observed radio emission pointed towards its origin being sub-relativistic outflows (e.g., \citealt{Alexander_2016, van_Velzen_2016, Krolik_2016,stein_2020}). Until recently, in all of the cases where TDE radio emission was detected, the interpretation of the data suggested that the outflow responsible for the radio emission was launched briefly after the stellar disruption (also in cases where radio emission is observed at late-times as in the case of the infrared TDE Arp 299B-AT1; \citealt{mattila_2018}). 

Conducting radio observations of TDEs not only early on, but also at late times (months and years after the stellar disruption) can provide key information on these events. For example, if a relativistic off-axis jet is launched, we may detect radio emission from this jet only at late times. The temporal and spectral evolution of the radio emission may reveal changes in the density structure of the CNM, energy injection processes, and otherwise new phenomena. Recently, \cite{Horesh_2020} reported a surprising discovery, the detection of a delayed radio flare from the TDE ASASSN-15oi (\citealt{Holoien_2016a}), several months after optical discovery with a second radio flare appearing years later. \cite{Horesh_2020} suggest, based on the temporal behaviour of the radio emission, that in the case of ASASSN-15oi the delayed radio flare is a result of a delayed launching of a high-velocity outflow. 

Here, we report the detection of another possibly delayed radio flare from an optically discovered TDE, namely iPTF\,16fnl \citep{Blagorodnova_2017}.  iPTF\,16fnl was discovered by the intermediate Palomar Transient Facility (iPTF) on 2016 August 29 (MJD\,=\,57629.0; and adopted as $T_0$ here) at $\alpha_{J2000} = 00^{\rm h} 29^{\rm m} 57^{\rm s}.04$, $\delta_{J2000} = +32\degr 53\arcmin  37\farcs5 $ in both the $g$ and $r$ optical bands, and was soon after spectroscopicaly classified as a TDE \citep{Blagorodnova_2017}. The location of iPTF\,16fnl is consistent with the center of its host galaxy at a redshift of $z=0.016328$ ($66$\,Mpc), making it one of the nearest optically discovered TDEs. \cite{Blagorodnova_2017} concluded, based on data collected as part of an extensive optical observational campaign, that iPTF\,16fnl is an atypical TDE, with a rather faint optical peak luminosity and a rapidly declining optical emission. They also reported several observations in the radio that resulted in non-detections. 

Here, we present an extended radio data set for iPTF\,16fnl and report the onset of late-time ($\sim 5$\,months after optical discovery) radio emission, that appeared after the initial radio non-detections reported by \cite{Blagorodnova_2017}. In \S2 we provide the full details of the radio data we obtained. We then discuss the radio emission detected at late time in \S3 and discuss its possible origin in \S4. We summarize in \S5.

\section{Radio Observations}
\label{sec:radio_obs}

Radio observations of iPTF\,16fnl were first reported by \cite{Blagorodnova_2017}. Their reported observations include data from both the Karl G. Jansky Very Large Array (VLA) and the Arcminute Microklevin Imager - Large Array (AMI-LA; \citealt{zwart_2008,hickish_2018}) and were undertaken on timescales of upto $\approx 55$\,days after optical discovery. \cite{Blagorodnova_2017} found that all of these relatively early observations resulted in non-detections. Here, we report additional VLA and AMI observations, undertaken at $63\textendash 417$\,days after optical discovery, and also report the detection of radio emission in some of these late-time observations.

\subsection{The Karl G. Jansky Very Large Array}
\label{sebsec: VLA}

We observed iPTF\,16fnl in radio wavelengths with the VLA. The broadband continuum observations (under our DDT program VLA/16A-225; PI Horesh) were conducted using C-, and K- bands at three epochs (in various telescope array configurations). 

Flagging and calibrations of the data were performed using the automated VLA calibration pipeline available in the Common Astronomy Software Applications (CASA) package \citep{mcmullin_2007}. Our gain calibrator was J0029+3456, and the primary flux density calibrator was 3C48. We imaged two sub-bands of C-band (centered at $4.8$ and $7.4$\,GHz; when in the case of a non-detection we combined the sub-bands), and the full K-band (centered at $22$\,GHz), using the CASA task CLEAN in an interactive mode, and the image rms was calculated using the CASA task IMSTAT. While the first two observations show no emission above $3 \sigma$, the last observation showed a point source at the phase center of both bands. We fitted this point source with the CASA task IMFIT and estimate the error of the peak flux to be the image rms, the fitting error produced by CASA, and a $10\%$ calibration error, summed in quadrature. The measured peak flux and the upper limits are reported in Table \ref{tab:Radio_Observations}.

\subsection{The Arcminute Microkelvin Imager - Large Array}
\label{sebsec: AMI-LA}

Radio observations of the field of iPTF\,16fnl were conducted with the AMI-LA, a radio interferometer located at the Mullard Radio Astronomy Observatory (MRAO) near Cambridge, England. AMI-LA observes with 28 baselines which extend from $18$\, to $110$\,m, and is comprised of eight, $12.8$\,m antennas. It operates with a $5$\,GHz bandwidth around a central frequency of $15.5$\,GHz. This results in a synthesized beam of $\sim 30$\,arcsec.

\begin{figure}[]
    \centering
    \includegraphics[width=0.45\textwidth]{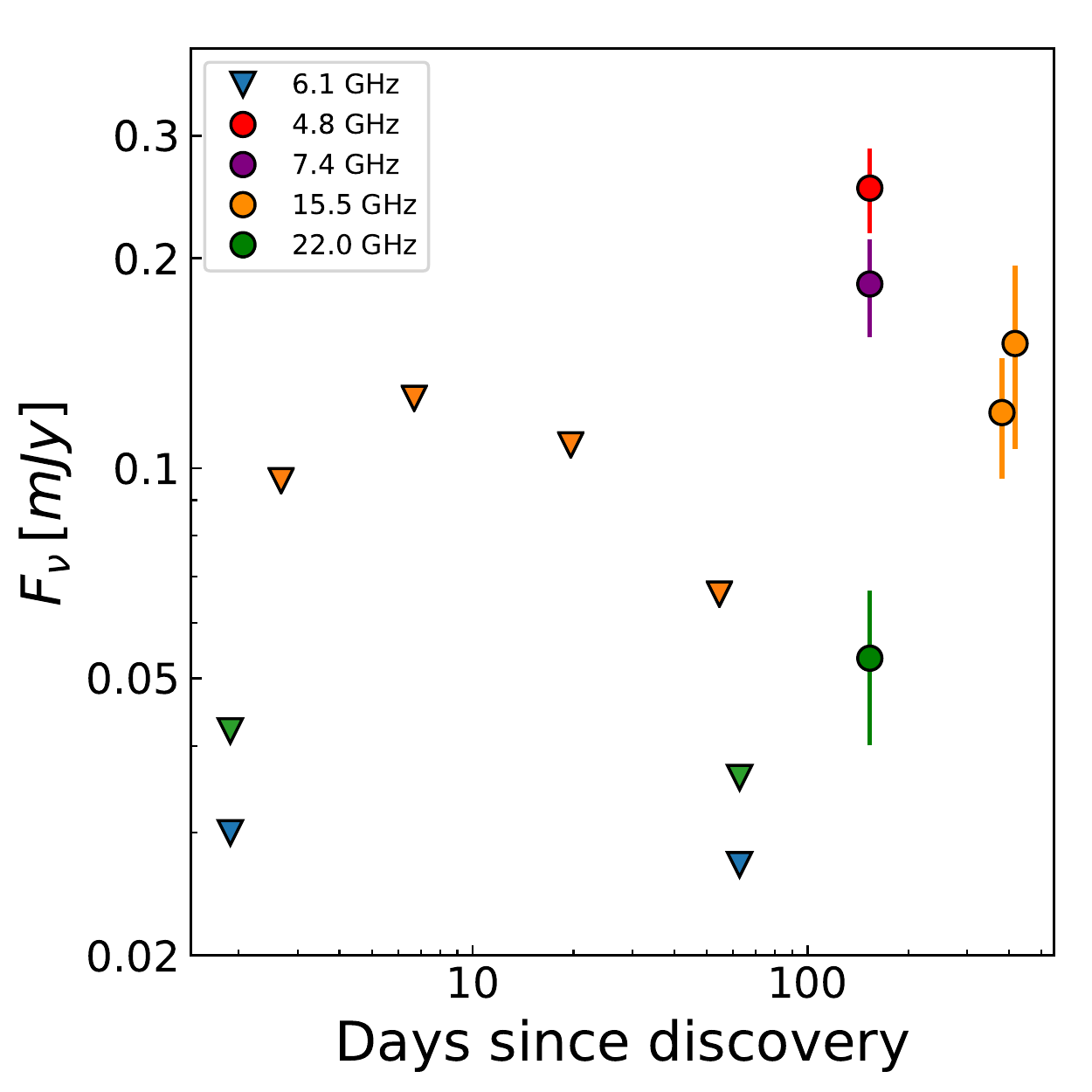}
    \caption{Radio light curves of iPTF16fnl as observed by the VLA and AMI-LA, and reported in \S\ref{sec:radio_obs}. Triangles represent $3\sigma$ upper limits while circles represent the flux density level of detected emission. The color of the triangles and circles refers to the central radio frequency.}
    \label{fig:light_curve}
\end{figure}

We first reduced our observations with $\tt{reduce \_ dc}$, a customized AMI-LA data reduction software package (e.g. \citealt{perrott_2013}) used for initial flagging and calibrations. Short interleaved observations of J0029+3456 were conducted to perform phase calibrations, and daily observations of 3C286 were used for absolute flux calibration. Additional flagging was performed using CASA. We imaged the field of iPTF\,16fnl using the CASA task CLEAN. Radio emission was detected from the source in the latter two observing epochs. The radio emission was measured with the CASA task IMFIT. The image rms was calculated using the CASA task IMSTAT, and we estimate the error of the peak flux to be the image rms, the fitting error produced by CASA, and $5\%$ calibration error, summed in quadrature. The measured peak flux and the upper limits are reported in Table \ref{tab:Radio_Observations} and plotted in Figure\,\ref{fig:light_curve}.

\section{Detection of Late-time Radio Emission}
\label{sec:results}

Our VLA observations show that up to $\approx 2$\,months after stellar disruption, no radio emission has been detected both at $6.1$\,GHz ($F_{\nu} \lesssim 0.03\,{\rm mJy}$) and $22$\,GHz ($F_{\nu} \lesssim 0.04\,{\rm mJy}$). Our AMI observations up to $\Delta t = 55$\,days, also resulted in non-detections with the deepest upper limit at $15.5$\,GHz obtained on $\Delta t = 55$\,days ($F_{\nu} \lesssim 0.07\,{\rm mJy}$). Radio emission, however, was detected in the third epoch of the VLA observations, at $\Delta t = 153$\,days after optical discovery. The radio emission seems to be optically thin ($F_{\nu}\propto \nu^{\alpha}$, with a spectral index of $\alpha \approx -1$) in the frequency range of $5 - 22$\,GHz. The difference in flux density levels between the latest $5$\,GHz non-detection and the detected emission on $\Delta t = 153$\,days, requires a temporal evolution as steep as $F_{\nu} \propto t^{2.5}$ at that frequency (in case radio emission is generated early on at this frequency and increase smoothly as a power-law function). Subsequent AMI observations at $\Delta t = 381$ and $\Delta t = 417$\,days revealed radio emission at $15.5$\,GHz which seems to be slowly increasing. The combined VLA and AMI data suggest that following the initial radio detection, the optically thin emission is increasing at a relatively slow rate (slower than a linear temporal evolution). We next examine our findings in light of theoretical models for TDE radio emission. 

\startlongtable
\begin{deluxetable}{cccc}
\tablecaption{iPTF\,16fnl - radio observations.\label{tab:Radio_Observations}}
\tablehead{
\colhead{$\Delta t$} & \colhead{$\nu$} & \colhead{$F_{\nu}$} & \colhead{Telescope} \\
\colhead{$[\textrm{Days}]$} & \colhead{$[\textrm{GHz}]$} & \colhead{$[\textrm{mJy}]$} &
}
\startdata
$1.89$ & $6.1$ & $< 0.030$ &  VLA:B \\ [0.5ex]
$1.89$ & $22$ & $< 0.042$ &  VLA:B \\ [0.5ex]
$2.68$ & $15.5$ & $< 0.096$ &  AMI-LA \\ [0.5ex]
$6.69$ & $15.5$ & $< 0.126$ &  AMI-LA \\ [0.5ex]
$19.6$ & $15.5$ & $< 0.108$ &  AMI-LA \\ [0.5ex]
$54.5$ & $15.5$ & $< 0.066$ &  AMI-LA \\ [0.5ex]
$62.7$ & $6.1$ & $< 0.027$ & VLA:A \\ [0.5ex]
$62.7$ & $22$ & $< 0.036$ & VLA:A \\ [0.5ex]
$153$ & $4.8$ & $0.252 \pm 0.034$ &  VLA:$\rm{A=>D}$ \\ [0.5ex]
$153$ & $7.4$ & $0.184 \pm 0.029$ &  VLA:$\rm{A=>D}$ \\ [0.5ex]
$153$ & $22$ & $0.053 \pm 0.013$ &  VLA:$\rm{A=>D}$ \\ [0.5ex]
$381$ & $15.5$ & $0.120 \pm 0.024$ &  AMI-LA \\ [0.5ex]
$417$ & $15.5$ & $0.151 \pm 0.045$ &  AMI-LA \\ [0.5ex]
\enddata
\tablecomments{A summary of the radio observations of iPTF\,16fnl. $\Delta t$ is the time since optical discovery ($T_0$). $\nu$ is the observed frequency in GHz. Flux density upper limits are $3 \, \sigma$ of the image. For the VLA, the letters appearing after the VLA telescope name represent the name of the VLA telescope array configuration.}
\end{deluxetable}

\section{THE ORIGIN OF THE RADIO EMISSION}
\label{sec:origin}

Theoretical models of the radio emission from TDEs explain the origin of the emission as an interaction of an outflow ejected promptly after stellar disruption with the SMBH CNM. The outflow can be either a relativistic jet (observed on- or off-axis; \citealt{giannios_2011,metzger_2012,generozov_2017}) or a sub-relativistic outflow (e.g. \citealt{Krolik_2016}). In both cases, the interaction of the outflow with the CNM leads to a shockwave in the CNM which in turn accelerates free electrons to relativistic velocities and enhances magnetic fields, and thus leads to the onset of synchrotron emission (e.g., \citealt{chevalier_1982,sari_1998}). Depending on the properties of the shockwave and the CNM, the synchrotron radio emission can also be absorbed below a certain frequency and thus the radio broadband spectrum may consist of both an optically-thick and -thin emission. Both parts of the radio spectrum are expected to follow a certain temporal evolution according to the nature of the CNM interaction. 

\subsection{A Sub-Relativistic Outflow}

The full description of the broadband (optically thick and thin) synchrotron emission from the interaction of a sub-relativistic outflow with the CNM is described by \cite{chevalier_1998}. In this case, the temporal evolution (see also \citealt{Krolik_2016}) of the optically thick emission is $F_{\nu,{\rm thick}}\propto t^{(8+k)/4}$ , and of the optically thin emission is  $F_{\nu ,{\rm thin}}\propto t^{(12-8k)/4}$, where $k$ is the power-law index of the CNM density profile ($\rho_{\rm CNM} \propto r^{-k}$). We next attempt to fit this model to the full observed dataset. 

\begin{figure}[]
    \centering
    \includegraphics[width=0.45\textwidth]{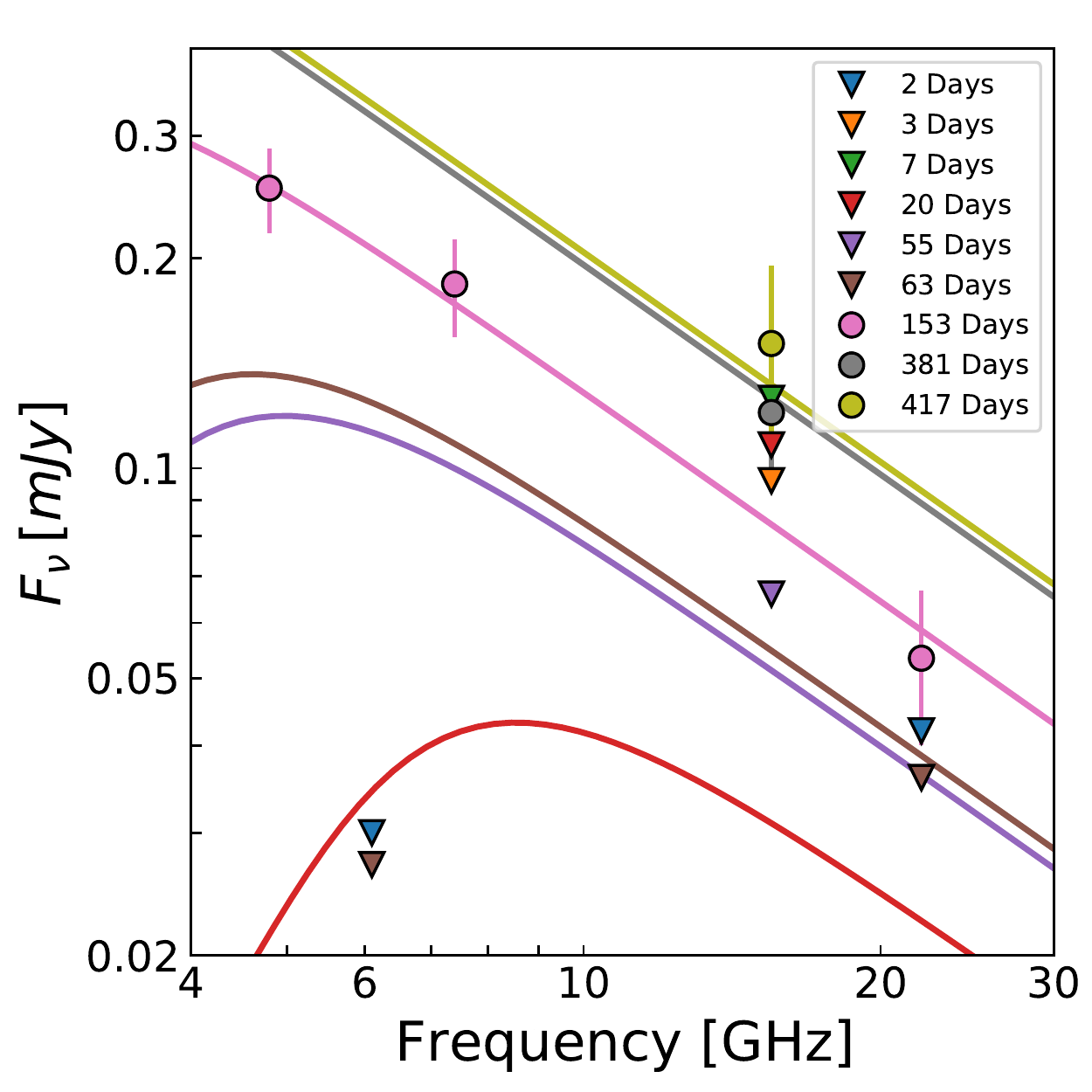}
    \caption{A fitted model for the radio emission originating from an outflow launched promptly after the stellar disruption and interacting with a CNM consisting of a single power-law density structure. This model can account for the VLA and AMI detected emission. However, the upper limit obtained at $63$\,days (brown triangle) strongly disagrees with the model. Triangles represent $3\sigma$ upper limits while circles represent the flux density level of detected emission.}
    \label{fig:t0_fit1}
\end{figure}

\begin{figure}[]
    \centering
    \includegraphics[width=0.45\textwidth]{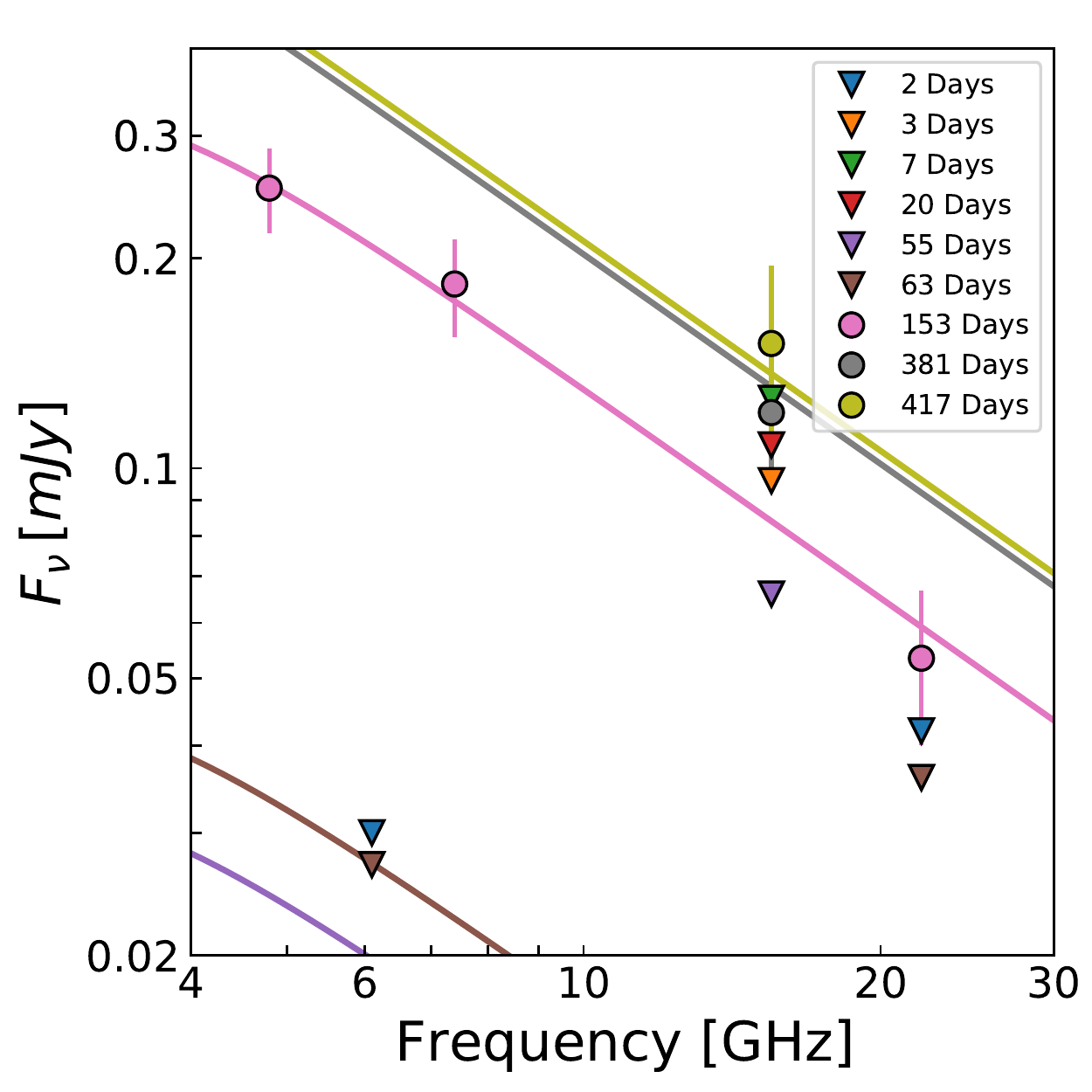}
    \caption{A fitted model for the radio emission originating from an outflow launched promptly after the stellar disruption and interacting with a CNM consisting of a varying two power-law density structure. In contrast to the single power-law CNM model presented in Figure~\ref{fig:t0_fit1}, this model is consistent with both the radio detections and non-detections. However, this model is somewhat at odds with the CNM profiles that are usually found in other TDEs (see text for details). Triangles represent $3\sigma$ upper limits while circles represent the flux density level of detected emission.}
    \label{fig:t0_fit2}
\end{figure}

We fail to find a model which is consistent with a sub-relativistic outflow launched promptly after stellar disruption into a CNM characterized by a single power-law density structure (as suggested, for example, by \citealt{Alexander_2016} and \citealt{Krolik_2016} for explaining the radio emission observed in ASASSN-14li). A temporal behaviour of $F_{\nu} \propto t^{5/2}$ at lower GHz frequencies is expected while the emission is optically thick (synchrotron self-absorbed) and originating from a shockwave traveling in a CNM with a density profile of $\rho\propto r^{-2}$. This temporal behaviour cannot account for the combination of the observed late-time radio emission with the early non-detections, at both $5$\,GHz and $22$\,GHz. While the best fit for a single power-law CNM model consists of a CNM density profile of $\rho \propto r^{-1.27}$, there is a strong disagreement between this model and the observed upper limit at $63$\,days (see Figure~\ref{fig:t0_fit1}). Instead, considering a varying CNM density profile, it may be possible to explain the full set of observations (including the early upper limits) with a shockwave traveling with a velocity $v_{s}\approx 1.8\times 10^{4}\,{\rm km\,s}^{-1}$ in a rather shallow CNM density profile of $\rho \propto r^{-0.4}$ that becomes $\rho \propto r^{-1.26}$ (Figure~\ref{fig:t0_fit2}) at a larger radius. While the outflow is traveling in the shallower part of the CNM structure, the radio emission stays optically thick and below our detection limits for a prolonged time. This model includes a somewhat arbitrary choice of the time of the VLA detection as the time when the change in the CNM density structure occurs. Note, however, that in several cases where CNM density profiles were measured in the close vicinity of SMBH using TDE observations, the CNM is steeper than $\rho \propto r^{-2}$ (as steep as $\rho \propto r^{-2.5}$) and sometimes become shallower (to the level of $\rho \propto r^{-1}$) only far away from the SMBH (\citealt{Alexander_2020}), in contrast to the fitted model here. Yet another possibility is that there is a CNM cavity in the close vicinity of the SMBH. In such a case radio emission will be generated only later on when an outflow (launched early on) finally reaches the CNM and interaction ensues. However, this type of a CNM structure with a cavity is unlike CNM structures observed around other SMBHs, and an explanation of how such a cavity can form is needed.

\begin{figure}[!ht]
    \centering
    \includegraphics[width=0.45\textwidth]{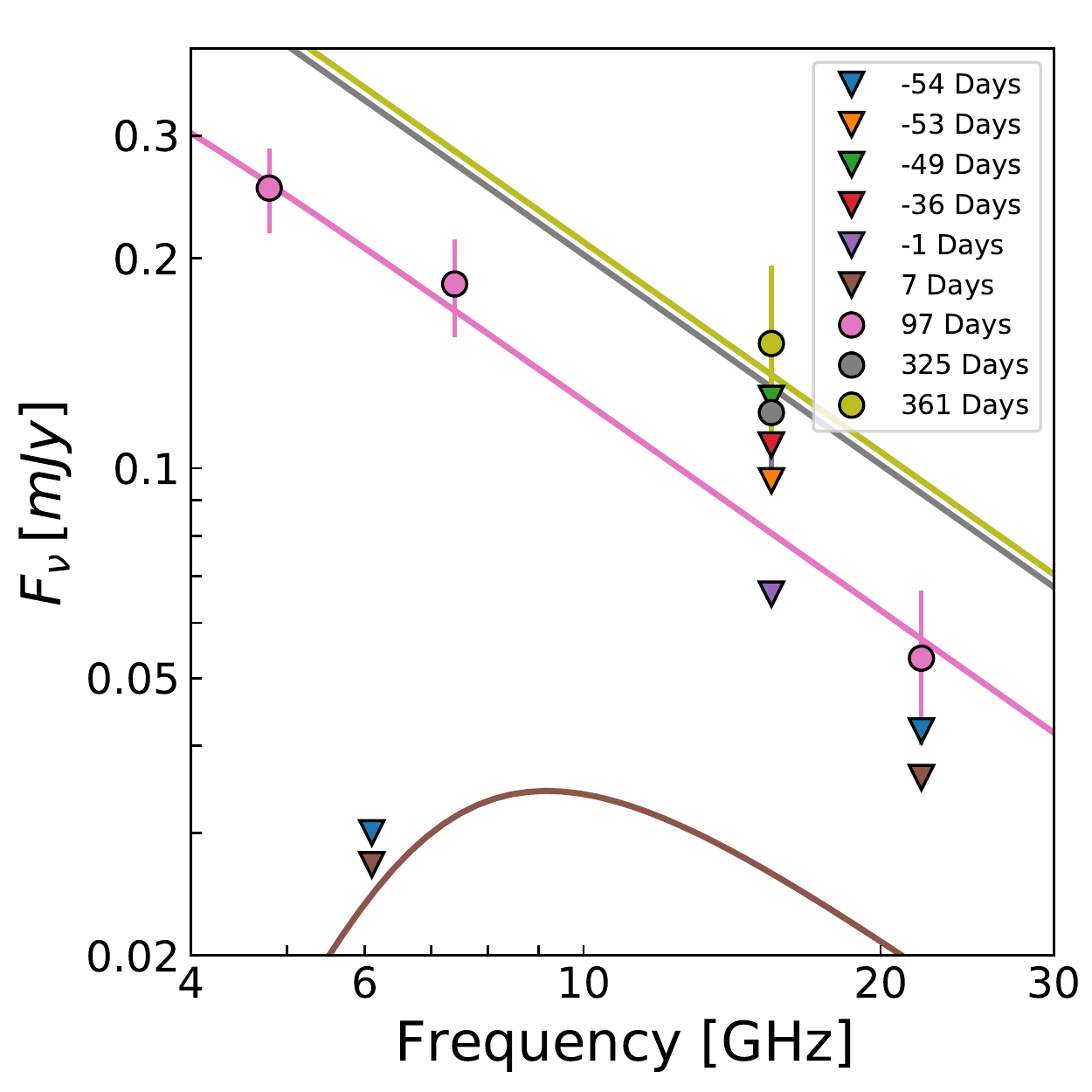}
    \caption{A model for the radio emission in the case where the emission is originating from the interaction of an outflow, launched only at late times ($\Delta t=56$\,days), with the CNM. In this case, the early non-detections can be explained by the fact that simply no significant radio emission was generated due to the lack of outflow-CNM interaction. An alternative to this model is the presence of a CNM cavity in the close vicinity to the SMBH, which can also result in a late-time onset of outflow-CNM interaction even if the outflow was launched early on.}
    \label{fig:t56_fit}
\end{figure}

The atypical CNM structures required to explain the delayed radio flare in a scenario when an outflow was launched briefly after stellar disruption, raises another possibility. Perhaps, we are simply observing a delayed outflow launching, similar to the suggested explanation for the delayed radio flare observed in ASASSN-15oi. Unfortunately, the available radio observations of iPTF\,16fnl are rather scarce compared to the more extended radio dataset available for ASASSN-15oi, thus rendering any detailed testing of this scenario more challenging. Still, assuming a delayed flare originating from a delayed ejection of a sub-relativistic outflow at time $t_{\rm delay}$, we find a possible solution for $t_{\rm delay}\geq 56$\,days. In this latter case, a single shockwave is traveling at a velocity of $v_{s}\approx 4.7\times 10^{4}$\,km\,s$^{-1}$ into a CNM with a density profile of $\rho\propto r^{-1.3}$ (see Figure~\ref{fig:t56_fit}). 

\subsection{A relativistic outflow}

A key past example of a relativistic TDE which exhibited radio emission is {\it Swift}\,J1644+57 \citep{Zauderer_2011,Berger_2012}. In this case the radio emission had high radio peak luminosity of $>10^{41}$\,erg\,s$^{-1}$. \cite{metzger_2012}, for example, model the radio emission in this event as an interaction of a relativistic jet with the CNM, a model which includes predictions for both the emission rise time and for its peak luminosity. Adopting this model, and assuming that the emission we observe in the case of iPTF\,16fnl at $\Delta t = 153$\,days is of the order of the radio peak emission, suggests that it is improbable that the observed radio emission is a result of an on-axis relativistic jet. The observed radio luminosity is rather weak ($<10^{37}$\,erg\,s$^{-1}$), compared to {\it Swift}\,J1644+57, and to accommodate an on-axis relativistic jet model, an extremely low CNM density is required (at a level of $\approx 10^{-3}$\,cm$^{-3}$). 

To test for the possibility that the observed radio emission is consistent with an off-axis jet model we attempt to fit a numerical model to the data using the BoxFIT code (\citealt{van_eerten_2012}). In the fitting process, we require that the flux will be below the non-detection level at the early observing epochs and use the full data set both from the VLA and the AMI telescopes. The BoxFIT code did not converge, and a suitable set of parameters that can fit the full set of observations was not found.

\pagebreak
\section{Conclusions and Summary}

Here, we reported the radio observations of the tidal disruption event iPTF\,16fnl from early times (a few days after optical discovery) until late times (more than a year later). Early on, no radio emission was detected at both $6$\, and $22$\,GHz at $\Delta t = 2$\ and $\Delta t = 63$\,days after optical discovery. The onset of an optically thin weak radio emission was discovered only at $\Delta t = 153$\,days after optical discovery. This late-time emission is observed to be rising with time in subsequent AMI observation at $15.5$\,GHz. Slowly rising weak radio emission in past TDE events such as ASASSN-14li \citep{Alexander_2016,Krolik_2016} has been explained by a sub-relativistic outflow launched soon after the stellar disruption and interacting with the CNM. Such an interaction drives a shockwave which in turn leads to synchrotron emission being emitted. In the case of iPTF\,16fnl, while the sequence of radio non-detections, combined with the observed radio emission later on and its temporal evolution, can be explained under this latter model, it requires an atypical CNM density structure (compared to CNM density structures observed in past TDEs), that combines both a spatially varying density profile in the close vicinity of the SMBH and a rather shallow profile. The odd properties of the CNM required for the sub-relativistic outflow model may suggest that some other process is at play. 

We next consider an alternative explanation. 
It is possible that these early non-detections suggest that no radio emission has been generated via interaction of an outflow with a CNM at early times. Thus, the discovery of the late-time onset of the radio emission may point to the onset of late-time interaction. If this is the case, then this can be due to having a CNM cavity around the SMBH out to a large distance or due to a late-time delayed outflow ejection. The former scenario requires some explanation on how such a cavity can form, while the latter is a scenario similar to the one that was recently suggested for explaining the delayed radio flares observed in ASASSN-15oi \citep{Horesh_2020}. 
The possibility that the late-time radio emission in iPTF\,16fnl is indeed due to a delayed outflow ejection suggests that ASASSN-15oi-like cases may be more common\footnote{Only three TDEs (including iPTF\,16fnl but excluding ASASSN-15oi) out of the $23$ events, which radio non-detections are summarized in \cite{Alexander_2020}, have both early (within a month after disruption) and late time ($\gtrsim 6$\,months) radio observations. Out of these three events, one (iPTF\,16fnl, given the additional observations presented here) exhibited late-time onset of radio emission.}. If this turns out to be the case, then future observations of TDEs in the radio (and in other wavelengths, especially X-rays) should also be carried out at late times, regardless of whether any emission has been detected early on. These late-time observations will play a key role in unveiling the nature of this newly discovered late-time phase in TDEs.

Uncovering the details of these possible delayed outflow ejections and their occurrence rate in TDEs may provide insight into accretion physics. In TDEs, delayed outflow ejections may be a result of a transition in accretion state and thus may provide a key piece of the puzzle of understanding what happens to the bound stellar debris from disruption until a much later time. Moreover, the knowledge gained from studying these late-time processes in TDEs may be applied to the study of active galactic nuclei (AGNs) and other accreting flaring systems. Thus, newly open questions, such as whether TDEs follow the patterns of disc-jet coupling established for X-ray binaries \citep{fender_2004} and likely also apply to ‘normal’ AGNs \citep{fernandez_2021}, can be explored.

\section*{Acknowledgements} 
 We thank the anonymous referee for their comments and suggestions. A.H. is grateful for the support by the I-Core Program of the Planning and Budgeting Committee and the Israel Science Foundation, and support by ISF grant 647/18. This research was supported by Grant No. 2018154 from the United States-Israel Binational Science Foundation (BSF). 
The authors thank the NRAO staff for approving and scheduling the VLA observations. The National Radio Astronomy Observatory is a facility of the National Science Foundation operated under cooperative agreement by Associated Universities, Inc. We thank the staff of the Mullard Radio Astronomy Observatory
for their assistance in the commissioning, maintenance and operation of
AMI, which is supported by the Universities of Cambridge and Oxford.

\bibliography{TDE.bib}


\end{document}